\documentclass[12pt,english]{article}
\usepackage[T1]{fontenc}
\usepackage[latin9]{inputenc}
\usepackage{geometry}
\usepackage{mathrsfs}
\usepackage{amsmath}
\usepackage{amssymb}
\usepackage{graphicx}
\usepackage{babel}
\usepackage[unicode=true,pdfusetitle,
 bookmarks=true,bookmarksnumbered=false,bookmarksopen=false,
 breaklinks=false,pdfborder={0 0 1},backref=false,colorlinks=false]
 {hyperref}

\begin{document}
\title{The ABC of scale invariance at the level of action integrals, and the software tool Kanon}
\author{R. Dengler\thanks{ORCID: 0000-0001-6706-8550}}
\maketitle

\begin{abstract}
A central and common aspect of renormalizable field theories is scale
invariance of the action integral. This note introduces the software
tool \emph{Kanon}\footnote{Available at Sourceforge,
\href{https://sourceforge.net/projects/kanon/}{https://sourceforge.net/projects/kanon/}},
which allows to assemble arbitrary action integrals interactively, and to determine
their critical dimension and scale invariance. The tool contains more
than 60 well-known  models with comments and references to the literature. 
\end{abstract}

\medskip{}
\tableofcontents{}
\medskip{}

\section{Introduction }
The field theoretic renormalization group has applications throughout
physics, but the huge number of field theories, their different forms,
as well as different regularization and renormalization schemes may
appear confusing.

It is all the more important to understand scale invariance at the
lowest order, scale invariance at the level of the action integral.
In a way and to some extent, a field theory is understood with the
scale invariance of its action integral. The perturbative renormalization
in principle is pure formalism and only adds some (often small) ``anomalous''
dimensions or possibly logarithmic factors.

The degrees of freedom in the renormalization procedure (which terms
need a coupling constant, which terms can be kept fixed, which scale
factors are required, ...) already appear at the level of the action
integral. This note describes the mathematics and physics of the algorithms
used in the \emph{Kanon} tool. Although simple in principle, this
may be useful to know also without the tool.

\section{Scale invariance of action integrals at the critical dimension}

Scale invariance of the action integral is a central and common aspect
of most renormalizable field theories. Formalizing scale invariance
considerations requires some conventions and some linear algebra.
An example is the Ginzburg-Landau-Wilson type action integral
\begin{equation}
S_{1}=\int\mathrm{d}^{d}x\left(a\boldsymbol{\varphi}\nabla^{2}\boldsymbol{\varphi}+u\left(\boldsymbol{\varphi}^{2}\right)^{2}+g\boldsymbol{\varphi}\nabla^{4}\boldsymbol{\varphi}+w\left(\boldsymbol{\varphi}^{2}\right)^{3}+v\boldsymbol{\varphi}^{2}\nabla^{2}\boldsymbol{\varphi}^{2}+r\boldsymbol{\varphi}^{2}+\boldsymbol{h}\cdot\boldsymbol{\varphi}\right)\label{eq:ActionGLW}
\end{equation}
describing several universality classes.
The action integral $S_{1}$ like most other ones is local in real space,
but it is customary to always talk of dimensions
of wave vectors $k$ (and frequencies) instead of coordinates $x$
(and time) in the renormalization group context. The terms of an action
integral are equivalent to monomials of wavevectors (or frequencies)
and fields because of scaling equivalences like $x\sim k^{-1}$, $\nabla\sim k$,
$\int\mathrm{d}^{d}x\sim k^{-d}$, $\int\mathrm{d}y\sim k_{y}^{-1}$,$\mathrm{\int d}t\sim\omega^{-1}$
and $\delta\left(y\right)\sim y^{-1}$. It generally suffices to have
one $d$-dimensional space apart from possibly other one-dimensional
spaces.

An action integral can be scale invariant under a rescaling
\begin{align*}
k_{m} & \rightarrow k_{m}b^{\left[k_{m}\right]},\\
\varphi_{m} & \rightarrow\varphi_{m}b^{\left[\varphi_{m}\right]}
\end{align*}
of coordinates $k$ and fields $\varphi$ with an arbitrary scaling
factor $b\neq0$ and constant scaling dimensions $\boldsymbol{N}=$
$\left\{ \left[k_{m}\right],\left[\varphi_{m}\right]\right\} $. The
scaling dimension corresponding to the $d$-dimensional space, $\left[k_{1}\right]=-\left[x_{1}\right]=1$,
is one by definition. In the language of dimensional analysis
this means that the scaling dimensions $\boldsymbol{N}$ count wave
vector factors. 

In general, there are several types of wavevectors $k_{m}$ (or coordinates)
and several types of fields $\varphi_{m}$, but the cartesian components
of an isotropic (or relativistically invariant) space all scale with
the same scaling dimension and need not be distinguished. Likewise,
the components of a tensor- or spinor-valued field all scale with the same scaling
dimension, and count as one field as far as scaling is concerned.

This leads to the concept of a \emph{model order} $M$, the sum of
the number of coordinates and fields with independent scaling dimensions.
The model order of the action integral (\ref{eq:ActionGLW}) is $2$,
there is one field and one coordinate.

Scale invariance of the $n$-th monomial of an action integral leads
to a linear equation
\begin{equation}
\sum_{m=1}^{M}E_{n,m}N_{m}=0,\label{eq:ScaleInvariance_Homogeneous}
\end{equation}
for the scaling dimensions, where the \emph{exponent matrix} $E$
is integer-valued, and contains a row for each monomial of the action
integral, a column for each coordinate and a column for each field. 

The exponent matrix for the action integral (\ref{eq:ActionGLW})
reads
\begin{equation}
E=\left(\begin{array}{cc}
2-d & 2\\
-d & 4\\
4-d & 2\\
-d & 6\\
2-d & 4\\
-d & 2\\
-d & 1
\end{array}\right).\label{eq:E_GLW}
\end{equation}
Eq.(\ref{eq:ScaleInvariance_Homogeneous}) contains $M$ unknowns:
$M-1$ scaling dimensions $\boldsymbol{N}$ and the space dimension
$d$. This means that $M$ rows of eq.(\ref{eq:ScaleInvariance_Homogeneous})
or $M$ monomials of the action integral suffice to determine $\boldsymbol{N}$
and $d$.

By arranging the monomials of the action in appropriate order, it
always is possible to use the first $M$ rows of $E$, which form
a square matrix $E_{\square}$ (that is, the first $M$ monomials
of the action integral, if there is any solution at all). A nontrivial
solution of eq.(\ref{eq:ScaleInvariance_Homogeneous}) then only exists
for $\det E_{\square}=0$, and this equation determines the critical
dimension $d=d_{c}.$ Different pairs ($M-$tupels) of rows from $E$
from Gl.(\ref{eq:E_GLW}) lead to different scale invariant field
theories. Their critical dimensions are $4$ (standard $\varphi^{4}$-model),
$3$ (tricritical point), $8$ (Lifshitz point) and $6$ (tricritical
Lifshitz point).

Once $M$ rows or monomials are selected and the critical dimension
$d_{c}$ and the scaling dimensions $\boldsymbol{N}=\left\{ \left[k_{m}\right],\left[\varphi_{m}\right]\right\} $
are determined, no degrees of freedom remain for the other rows of
eq.(\ref{eq:ScaleInvariance_Homogeneous}) (monomials of the action
integral). These monomials may be consistent with the scaling (marginal)
or inconsistent (relevant or irrelevant).

There is a simple geometric picture for the scale invariance of an
action in an $M$-dimensional exponent space. The exponents $E_{n,m}$
of a monomial with index $n$ are points in this space with integer
coordinates. The scaling dimensions $\boldsymbol{N}$ define a vector
in the exponent space, and equation (\ref{eq:ScaleInvariance_Homogeneous})
says, that row $n$ of the exponent matrix (the $n$th monomial of
the action integral) must be orthogonal to the vector $\boldsymbol{N}$.
In other words, the monomials of a scale invariant field theory lie
on a hyperplane through the origin perpendicular to $\boldsymbol{N}$.

The exponents $E_{n,m}$ contain the a priori unknown dimension $d$,
but according to $E_{n,m}=E_{n,m}^{\left(0\right)}-\delta_{m,1}D_{n}d$
the dimension $d$ only enters as a translation in the $k_{1}$ direction.
In most cases all monomials contain exactly one $d$-dimensional integral,
and $D_{n}=1$. For $E_{n,m}^{\left(0\right)}$ scale invariance then
requires
\begin{equation}
\sum_{m=1}^{M}E_{n,m}^{\left(0\right)}N_{m}=d,\label{eq:ScaleInvariance_Inhomgeneous}
\end{equation}
and the critical dimension can be read off at the intersection of
the hyperplane spanned by the monomials $E_{n,m}^{\left(0\right)}$ with the $k_{1}$-axis.

\begin{figure}
\includegraphics[scale=0.3]{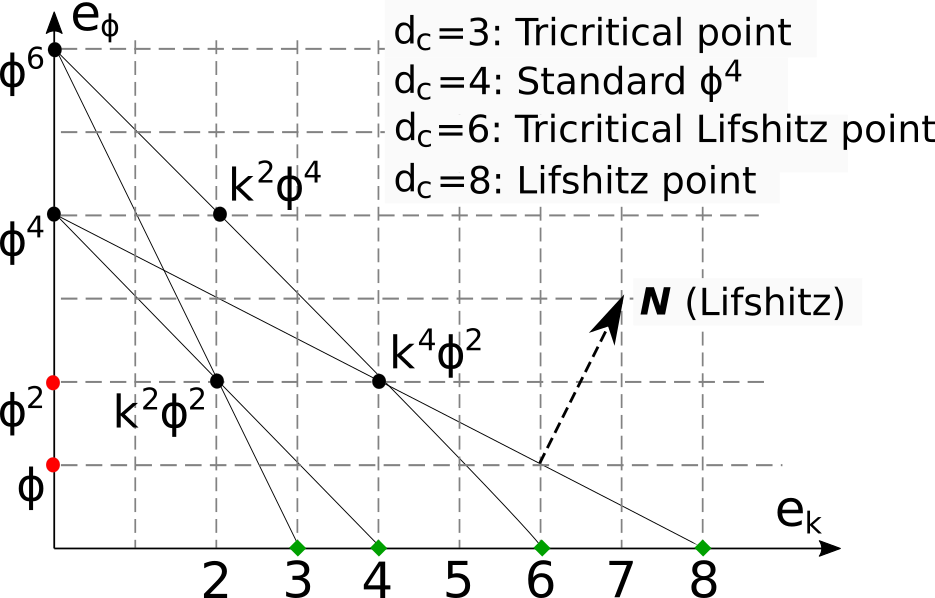}
\caption{\label{fig:Hyperplane}The exponent space for action integral (\ref{eq:ActionGLW})
with model order $M=2$. A hyperplane represents a model with a critical
point. Monomials on the upper right of a hyperplane are relevant for
large wavevectors and irrelevant for small wavevectors. Monomials
on the lower left of a hyperplane are relevant for small wavevectors
and irrelevant for large wavectors.}
\end{figure}
This is shown for action (\ref{eq:ActionGLW}) in fig.(\ref{fig:Hyperplane}).
Geometrically, equations (\ref{eq:ScaleInvariance_Inhomgeneous})
state that the exponents $E_{n,m}^{\left(0\right)}$ of marginal monomials
can be interpreted as points on a hyperplane in a model-order dimensional
exponent space. The hyperplane is orthogonal to the vector $\boldsymbol{N}$
of scaling dimensions. The critical dimensions $d_{c}$ together with
the scaling dimensions $\boldsymbol{N}$ are a (not unique) signature
for a scale invariant action integral. 

To summarize, a scale invariance of an action integral requires selecting
\emph{model oder} monomials of the action integral. These monomials
define a hyperplane in exponent space and determine the critical dimension
and the scaling dimensions. It then may turn out that more (or even
an infinite number of) monomials are consistent with the scale invariance. 

\section{Scaling below or above the critical dimension}

The algebra decribed above must be extended to $d\neq d_{c}$ when
the actual space dimension $d$ differs from the critical dimension
$d_{c}$ or when dimensional regularization is to be used. Scale invariance
is possible for $d\neq d_{c}$ if a coupling constant is assigned
to \emph{one} of the first $M$ monomials of the action integral.
Which of the first $M$ monomials is assigned a coupling constant
is more or less is a matter of taste.\footnote{It is possible to assign a coupling constant to several of the first
model order monomials, but the solution then is not unique. In the
\emph{Kanon} tool a unique solution can be enforced by adding a constraint
monomial (unphysical) to the action integral.}

The coupling constant then also gets a wave vector dimension and takes
part in the scaling. The coupling constant adds a column to the exponent matrix and a
row to the vector of scaling dimensions $\boldsymbol{N}$, but the
first column of the exponent matrix and the value $N_{1}=1$ are known
and can be moved to the r.h.s. of eq.(\ref{eq:ScaleInvariance_Homogeneous}).
The $d$-dependent scaling dimensions then follow by inverting the
resulting $M\times M$ matrix. The wavevector dimension of the coupling
constant is of order $O\left(\epsilon\right)$, with $\epsilon=d_{c}-d$.

The resulting scale invariance of the action integral looks like a
renormalization group flow for the coupling constant. This flow of
coupling constants with a naive scaling dimension of order $O\left(\epsilon\right)$
is a part of the physical coupling constant flow, but has a physical
meaning only when combined with the contribution from perturbation
theory, which has the same order of magnitude.

No degrees of freedom are left once the $M\times M$-equation is solved,
and in general all monomials after the first $M$ monomials need a
coupling constant.\footnote{In some cases it turns out that there are relations between these
coupling constants. Examples are nonlinear gauge theories and mode
couplings in critical dynamics.}

The software tool \emph{Kanon} always determines the critical dimension
and the canonical scaling dimensions from the first $M$ monomials
of the action integral, and allows to shift monomials up and down
by drag and drop.

Shifting monomials with a coupling constant of order $O\left(\epsilon\right)$
(marginal, on the same hyperplane) corresponds to arbitrary assignments
in the renormalization scheme (which of the first $M$ vertex functions
are kept fixed, which one of them has a coupling constant). The canonical
dimensions in general change at order $O\left(\epsilon\right)$, but
the physical scaling dimensions remain the same.

In contrast, when monomials with a coupling constant of order $O\left(1\right)$
(relevant or irrelevant, on a different hyperplane) are shifted to
the first $M$ monomials, a different universality class (another
hyperplane, or none at all) gets selected. See also fig.(\ref{fig:Hyperplane}).

\section{Dimensionless fields}

The number of possible marginal monomials is finite as long as the
hyperplane is not parallel to some field axis, i.e., as long as all
scaling dimensions $\boldsymbol{N}$ differ from zero. It sometimes
occurs that some field $\varphi_{m}$ is dimensionless, $\left[\varphi_{m}\right]=0$.
Geometrically this means that the hyperplane is parallel to the field
exponent axis$,$ and that it intersects an unlimited number of lattice
points (fig.(\ref{fig:Hyperplane})). Multiplying any marginal monomial
of the action integral with any power of $\varphi_{m}$ generates
another marginal monomial, and in principle an infinite set of monomials
are marginal and must be taken into account.

The field theory and perturbative renormalization might still make
sense. Some monomials might be forbidden by symmetry and not generated
by perturbation theory. Or the perturbative expansion only involves
these monomials step by step, and an iterative procedure is possible.
But the situation always should be examined meticulously. Examples
of field theories with a dimensionless field are nonlinear sigma models,
the KPZ (Kardar, Parisi, Zhang) interface growth model, the Sine-Gordon
model and the Wilson-Cowan neural network model.

\section{Symmetries between fields}

Some action integrals are symmetric under an exchange of two fields.
An example is the Fermi action of beta decay\cite{Scadron1979}
\[
S_{\mathrm{F}}=\int\mathrm{d}^{d}x\bar{\psi}\left(\gamma\cdot\partial+m\right)\psi+\left(\bar{\psi}\gamma\left(1-\gamma_{5}\right)\psi\right)\cdot\left(\bar{\psi}\gamma\left(1-\gamma_{5}\right)\psi\right).
\]
The model order is three. The mass $m$ is irrelevant for large wavevectors,
and there remain two monomials in $S_{F}$ - not enough to determine
$d_{c}$, $\left[\psi\right]$ and $\left[\bar{\psi}\right]$. But
because of the symmetry $\psi\leftrightarrow\bar{\psi}$ it suggests
itself to require $\left[\psi\right]=\left[\bar{\psi}\right]$. This
amounts to using only one field in the dimensional analysis, or to
adding a constraint monomial $\bar{\psi}\psi^{-1}$ to $S_{F}$ (only
for the scaling analysis, without any integrals!). 

Any other constraint term like $\psi$ or $\bar{\psi}\psi^{3}$ not
symmetric under the $\psi\leftrightarrow$$\bar{\psi}$ symmetry also
does the job, even if this makes a field dimensionless. In fact, the
two fields always occur in pairs, and only the sum of the dimensions
$\left[\psi\right]+\left[\bar{\psi}\right]$ has physical meaning.

\section{Field theories in wavevector space}

In situations where a Fermi shell plays a role the field theory must
be written down in $k$-space. Examples are the Fermi liquid with
a four-point-interaction (superconductivity) and the Kondo effect.
The algorithm used in \emph{Kanon} for the dimensional analysis is
oblivious to whether a ``coordinate'' denotes a coordinate or a
wavevector. However, to get the correct sign for the scaling dimensions
requires to set a flag for the coordinate. And of course, dimensions
of fields and Fourier transforms of fields differ by the dimensions
of the Fourier transform integrals.

\section{Statistical physics and particle physics}

With respect to scale invariance and the renormalization group there
essentially is no difference between field theories of statistical
physics and of particle physics. In the latter case the action occurs
with a factor $i$ in the path integral, but this does not affect
the essence of the renormalization group formalism.

The actual difference is the perspective. In statistical physics a
system with atoms or molecules and usually a lattice is given, which
defines an ultraviolett cutoff (UV), and one is interested in the
behavior at decreasing wavevectors (long wavelengths, ``infrared'',
IR).

In partice physics the starting point is the physics at small wavevectors
(low energy), and the question is how this came about or what will
happen at high energies. Accordingly, one is interested in the behaviour
for growing wavevectors.

A primitive example is the action $S=\int\mathrm{d}^{d}x\left(\left(\nabla\varphi\right)^{2}+m\varphi^{2}\right)$.
The mass term $m\varphi^{2}$ dominates for small wavevectors (it
is relevant) in comparison to $\left(\nabla\varphi\right)^{2}$, and
this is what is of interest in statistical physics. In contrast, for
large wavevectors (particle physics) the mass term is negligible (irrelevant).
The difference only is the perspective.

\section{Connection with the field theoretic renormalization group}

Scale invariance of an action integral is a precondition for renormalizability,
and a scaling analysis always should be the first step of a renormalization
group calculation. The action integral must contain all types of marginal
interactions, in particular if such interactions are generated by
perturbation theory and/or are allowed by symmetry. If these preconditions
are met, then renormalizibility is the rule rather than the exception.

The naive scaling at the level of the action integral contains the
gist of scale invariance, but this exact classical symmetry is broken
when fluctuations are taken into account (an example of an ``anomaly'').

The actual scaling dimensions of the fields and coordinates and the
flow of the coupling constants only follow from a renormalization group calculation.
The field theoretic renormalization
group is perturbative in nature, and gives reliable results only for
small coupling constants. The corrections to the naive scaling then
also are small. However complicated the calculations are, there remain
a few typical possibilities.

\subsection{At the critical dimension}

In the vicinity of the trivial fixed point (with vanishing coupling
constants) the renormalization group adds small logarithmic corrections
to the naive scaling. If the fixed point is stable, then this in principle
is the complete answer.

Examples are quantum chromodynamics with asymptotic freedom (UV stable
trivial fixed point), or the standard $\varphi^{4}$-model in four
dimensions with an IR-stable fixed point, or the Uniaxial magnet with
dipolar interaction in three dimensions, also with a stable IR fixed
point.

If the fixed point is unstable, then one knows the physics near the
fixed point. But some coupling constant grows, and the ultimate behaviour
away from the fixed point cannot be described with the pertubative
renormalization group. Examples are quantum electrodynamics (UV unstable
fixed point), quantum chromodynamics (IR unstable fixed point, confinement)
and the Kondo effect also with an IR unstable fixed point. A Fermi
liquid corresponds to an IR stable fixed point in any space dimension,
but this fixed point is unstable under an attractive two-particle
interaction (ultimately leading to superconductivity).

\subsection{Below or above the critical dimension}

Below (or sometimes above) the critical dimension $d_{c}$ an expansion
parameter is $\epsilon=d_{c}-d$, and there may be fixed points of
the coupling constant flow equations of order $O\left(\epsilon\right)$. 

If there are stable fixed points for the coupling constants, then
there are $\epsilon$-expansions for critical exponents and other
quantities. This typically is the case in critical statics and critical
dynamics, and for systems of the reaction-diffusion type (e.g. percolation).

If one is more interested in numerical values instead of in exact
$\epsilon$-expansions, it also is possible to perform the renormalization
group calculations directly in the given dimension $d$, and to expand
quantities in the renormalized coupling constants (provided these
are small in $d$ dimensions).

\subsection{Non-renormalizable field theories}

Quantum field theories are not renormalizable above their critical
dimension.\cite{ZJ1997} Technically this means that there is a relevant interaction,
and that the degree of divergence of diagrams for a vertex increases
with the number of interactions. No finite number of subtractions
renders the vertex finite. Such field theories still can make sense
at tree level, as effective field theories. Examples are Fermi theory
of weak interactions and Einstein-Hilbert gravitation\cite{Scadron1979}, both with a
critical (spacetime) dimension two.

\section{Some special cases}

There are some peculiar cases of renormalizable field theories, where
a naive scaling analysis as described above can be misleading or somewhat
involved. 

\subsection{Sine Gordon model}

A monomial of the action integral can be strongly but \emph{dangerously}
irrelevant. This means that the monomial rapidly scales to zero according
to naive scaling analysis (it is not on the hyperplane), but must
be kept nevertheless. An example is the Sine-Gordon model\cite{ZJ1997} with action
\[
S_{\mathrm{sg}}=\int\mathrm{d}^{d}x\left\{ \left(\nabla\varphi\right)^{2}-\alpha\cos\varphi\right\} .
\]
One of several pecularities of the Sine-Gordon model is that one coupling
constant suffices for all powers of $\varphi^{2}$ in the $\cos\varphi$
series. The $\cos$-function can only play a role in this way if $\varphi$
is dimensionless. The first monomial then is marginal for $d=2$,
while the $\cos\varphi$ would be marginal for $d=0$, but is strongly
\emph{relevant} in two dimensions. 

A more standard version of the action contains an auxiliary field\footnote{The $4\pi$-periodicity
of $S_{\mathrm{sg}}'$ in $\varphi$ is reminiscent
of fermions and bosonization, and in fact $S_{\mathrm{sg}}$ can be
shown to be equivalent to the massive Thirring model.\cite{ZJ1997}} $\sigma$,
\[
S_{\mathrm{sg}}'=\int\mathrm{d}^{d}x\left\{ \left(\nabla\varphi\right)^{2}-\lambda\sigma\cos\tfrac{\varphi}{2}+\tfrac{1}{2\gamma}\sigma^{2}\right\} .
\]
Integrating out $\sigma$ from $e^{-S_{\mathrm{sg}}'}$ leads back
to $S_{\mathrm{sg}}$, but the monomials $\int\mathrm{d}^{d}x\sigma\varphi^{2n}$
now are marginal in two dimensions, the $\sigma^{2}$ is IR irrelevant.
However, the $\sigma^{2}$ must be kept to be able to integrate over
the auxiliary field $\sigma$. Calculations with action $S_{\mathrm{sg}}'$
are not simpler than calculations with $S_{\mathrm{sg}}$. But the
dimensional analysis is standard, and the first monomials of $S_{\mathrm{sg}}'$
as usual define a hyperplane in exponent space (fig.(1)).

\subsection{Kondo-effect}
A system can be inhomogeneous in space \emph{and} scale invariant.
An example is the Kondo effect\cite{Kondo1964} with action integral
\begin{align*}
S_{\mathrm{Kondo}} & =\int d\tau\int\mathrm{d}^{d}k\int\mathrm{d}\Omega\bar{\psi}\left(\partial_{\tau}+k\right)\psi+\int\mathrm{d}\tau\bar{\varphi}\partial_{\tau}\varphi\\
 & \qquad+\int\mathrm{d}\tau\left(\left(\int\mathrm{d}^{d}k\int\mathrm{d}\Omega\bar{\psi}\right)\boldsymbol{\sigma}\left(\int\mathrm{d}^{d}k\int\mathrm{d}\Omega\psi\right)\right)\cdot\left(\bar{\varphi}\boldsymbol{\sigma}\varphi\right)
\end{align*}
The symbol $\tau$ denotes imaginary time, $\psi$ describes band
electrons, $\varphi$ describes the defect electron at the origin.
The action integral is written in $k$-space because of the Fermi
surface for the band electrons. Components $\Omega$ of wavevectors
parallel to the Fermi surface take not part in the scaling and only
label the degrees of freedom (there is no distinguished $\Omega_0$ around which to scale $\Omega$).
The symbol $k$ denotes the wavevector
component perpendicular to the Fermi surface, and only $d=1$ has
physical meaning. This is consistent with the critical dimension $d_{c}=1$
(this already indicates that the Kondo effect occurs in any space dimension\footnote{Except for Luttinger
liquids with $d_{\mathrm{space}}=1$.} $d_{\mathrm{space}}=d_{c}+\mathrm{dim}\left(\Omega\right)$).
The nonlinear monomial is the interaction energy between band electrons
and the defect electron, $\boldsymbol{\sigma}$ denotes the Pauli
spin matrices (note that $\int\mathrm{d}^{d}k\int\mathrm{d}\Omega\psi\left(\boldsymbol{k}\right)=\psi\left(\boldsymbol{x}=0\right)$).

The action integral $S_{\mathrm{Kondo}}$ looks complicated, but the
scaling analysis is pure formalism and accomplished in Kanon on the
fly. The Pauli matrices $\boldsymbol{\sigma}$ and $\int\mathrm{d}\Omega$
are dimensionless and take not part in the scaling. As usual, once
it is clear that $S_{\mathrm{Kondo}}$ is scale invariant, the way
is open for a renormalization group calculation.

The fields $\varphi$ und $\bar{\varphi}$ actually only depend on
imaginary time $\tau$ and thus are not fields in the usual sense.
To remedy this one could write $\varphi=\Phi\left(\boldsymbol{x}=0\right)=\int\mathrm{d}^{d}k\int\mathrm{d}\Omega\Phi\left(\boldsymbol{k}\right)$
with a field $\Phi$. The scaling analysis also is standard with this
notation. However, the two notations lead to hyperplanes with different
normal vectors, i.e. the signatures differ. An advantage of the formulation
with the $\Phi$-field might be, that insisting on using fields in
the usual sense leads to a unique normal vector (signature) for the
model.

\section{Classification of renormalizible field theories, signature}

As illustrated in fig.(1), renormalizable field theories are associated
with hyperplanes in exponent space. Monomials on the hyperplane are
marginal, monomials on one side are relevant, monomials on the other
side are irrelevant.

Selecting different $M$ monomials from a set of monomials often selects
different hyperplanes, and makes different monomials relevant, marginal
or irrelevant.

This suggests to use the hyperplanes as signature for a critical model.
A hyperplane contains at least model order linearly independent monomials.
It can contain more monomials, all marginal and possibly important
for the critical point. The correspondence is not unique, different
universality classes can have the same hyperplane. For instance, the
naive scaling analysis is the same for scalar und vector valued order
parameters.

The signature consists of the critical dimension and the normal vector
$\boldsymbol{N}=\left(\left[k_{m}\right];\left[\varphi_{m}\right]\right)$
of the hyperplane. The naive scaling dimensions $\left[k_{m}\right]$
and $\left[\varphi_{m}\left(\boldsymbol{x}\right)\right]$ can be
arranged in increasing order, except for the first element $\left[k_{1}\right]=1$
for the $d$-dimensional space, which plays a special role. The table
shows some examples

\begin{table}
\centering{}%
\begin{tabular}{|c|c|c|}
\hline 
Name & Critical dimension & Signature\tabularnewline
\hline 
\hline 
Directed percolation & 4 & $\left(1,2;2,2\right)$\tabularnewline
\hline 
Dynamical percolation & 6 & $\left(1,2;2,2,4,4\right)$\tabularnewline
\hline 
Model A of critical dynamics & 4 & $\left(1,2;1,3\right)$\tabularnewline
\hline 
Model J of critical dynamics & 6 & $\left(1,4;2,4\right)$\tabularnewline
\hline 
Gauge field theory & 4 & $\left(2;2,3,3\right)$\tabularnewline
\hline 
Fermi theory of weak interactions & 2 & $\left(2;1,1\right)$\tabularnewline
\hline 
Fermi liquid & 1 & $\left(2,2;1,1\right)$\tabularnewline
\hline 
\end{tabular}\caption{Signatures of some renormalizable field theories. ``Critical dimension''
here is the dimension of the $d$-dimensional space, not the some
of all coordinate (and time) dimensions.}
\end{table}

The vector element $N_{1}=1$ conveys no information, and it is possible
to multiply $\boldsymbol{N}$ with the smallest integer converting
all $N_{m}$ to integers. For action integrals formulated in wavevector
space (instead of real space) the fields are functions of wavevectors,
and \emph{Kanon} calculates the dimensions of these fields. But to
have an invariant signature one should use the dimensions of real
space fields in the signature, which differ by the dimensions of the
Fourier transform integrals. It also suggests itself to list coordinate
dimensions (before the semicolon) and field dimensions (after the
semicolon) in increasing order, except for the $d$-dimensional coordinate,
which is the first element.

An interesting question is: how many renormalizable field theories
are there for a given number of coordinates and a given number of
fields? Exponents of fields in an action integral must be positive
(the only known exception is conformal quantum mechanics).

Exponents of coordinates can be negative because of integrals (and
sporadically inverse Laplace operators), but normally there is only
one integral per coordinate (in real space).

If only a limited number of field factors and a limited number of
derivates is allowed in each monomial, then the renormalizable field
theories correspond to hyperplanes defined by $M$ monomials (points
with integer coordinates) in some cuboid (or simplex) in $M$-dimensional
exponent space (fig.(1)).

To find the hyperplanes defined by $M$ points with integer coordinates
in this cuboid and other points on each hyperplane is a diophantine
problem.

This problem can be attacked somewhat brute-force with the help of
a computer, but not all hyperplanes lead to action integrals of the
usual type. For instance, a hyperplane must contain monomials harmonic
in the fields defining propagators, and more heuristic rules can
be imposed. The mapping from a hyperplane to a renormalizable field
theory is not unique, but it often is not difficult to write down
plausible candidates. It would be of interest to classify all scale
invariant action integrals for a given number of coordinates and fields
and a given cuboid of exponents.

\subsection{Additional marginal monomials}

A related but simpler diophantine problem is to find all marginal
monomials (in some range) of a given scale invariant action integral.
This amounts to solving a linear diophantine equation. The structure of the
solutions of linear diophantine equations is known\cite{Bundschuh1991}. There are $M-1$ linearly independent
base vectors $e_{i}$, and the lattice $m_{0}+\sum n_{i}e_{i}$ with
a marginal monomial $m_{0}$ and integers $n_{i}$ represents all
marginal monomials.

Numerically, a simple brute-force search for additional marginal monomials
in a cuboid of exponents turns out to be more efficient. The advantage of this
approach is that the iterations run over exponents and that the iteration
range has direct meaning. \emph{Kanon} offers both algorithms.

In any case, one should check additional marginal monomials carefully.
If such a monomial is generated in perturbation theory, then it must
be taken into acount in renormalization group calculations.\cite{Dengler1985}

Many other marginal monomials can be ruled out by physical or formal
reasons. If this is not the case, then adding the monomial to the
action integral may convert it to another universality class. The
question then is, whether the model with the additional interaction
has a physical meaning.

\section{Summary}

The software tool Kanon offers a unified view on all types of renormalizable
field theories. Action integrals can be defined interactively by drag
and drop, and if an action integral turns out to be scale invariant,
then what remains to be done is the renormalization group calculation.

When using the tool it should become clear, that a renormalizable
field theory is associated with a hyperplane in an exponent space,
determined by model order monomials. It should become clear that dimensionless
fields are problematic, and it should become clear that it is to some
degree arbitrary which monomials are kept fixed in the renormalization
procedure, and which ones need a coupling constant.
There often are additional marginal monomials, which can be determined
in Kanon on the fly. Most of them can be ruled out by formal or physical reasons, but the other
ones should be examined carefully.

The dimensional analysis of an action integral often is presented
in the literature in concrete cases quite ad hoc with arguments of different
quality. Scale invariance, however, is a precise mathematical concept,
and having a tool for this purpose is useful for practical and didactic reasons.

\bigskip{}
\bigskip{}
\bigskip{}

\end{document}